# Leptonic Signals from off-shell Z Boson Pairs at Hadron Colliders

C. Zecher $^{a*}$, T. Matsuura $^{b\dagger}$, J.J. van der Bij $^c$

$^a$ Deutsches Elektronen Synchrotron DESY, Notkestraße 85,
D-22603 Hamburg

$^b$ II. Institut für Theoretische Physik, Universität Hamburg,
Luruper Chaussee 149, D-22761 Hamburg

$^c$ Fakultät für Physik, Hermann-Herder-Straße 3, D-79104 Freiburg i.Br.

## Abstract

We study the gluon fusion into pairs of off-shell Z bosons and their subsequent decay into charged lepton pairs at hadron colliders : $gg \to ZZ \to 4\ell^\pm$ ($\ell^\pm$ : charged lepton). Throughout this paper we do not restrict the intermediate state Z bosons to the narrow width approximation but allow for arbitrary invariant masses. We compare the strength of this process with the known leading order results for $q\bar{q} \to ZZ \to 4\ell^\pm$ and for $gg \to H \to ZZ \to 4\ell^\pm$. At LHC energies ($\sqrt{s} = 14$ TeV) the contribution from the gluon fusion background is around 20 % of the contribution from quark-antiquark annihilation. These two processes do not form a severe irreducible background to the Higgs signal. At Higgs masses below 120 GeV the final state interference for the decay channel $H \to ZZ \to 4\mu^\pm$ is increasingly constructive. This has no effect on the Higgs search as in this mass region the signal remains too small. One can extend the intermediate mass Higgs search via off-shell Z boson pairs at the LHC down to about 130 GeV Higgs mass. However a careful study of the reducible background is needed for definitive conclusions.

---

$^*$email address: zecher@ips103.desy.de
$^\dagger$Now at Shell Research, Rijswijk, The Netherlands

# 1. Introduction

One of the main tasks of current and future colliders is to reveal the origin of the electroweak symmetry breaking. This includes first of all the search for a standard model Higgs boson. A light Higgs boson ($m_H < m_Z$, $m_H$ Higgs mass, $m_Z$ Z mass ) can be found at LEP200 assuming a sufficiently high integrated luminosity of order 0.5...10 fb$^{-1}$ [1, 2, 3]. Here the discovery limit is not dominated by a vanishing Higgs production rate but by the problem to distinguish a via $H \to b\bar{b}$ decaying Higgs boson with $m_H$ around $m_Z$ from a hadronically decaying Z boson. For an intermediate mass Higgs boson ( $m_Z < m_H < 2m_Z$ ) an $e^+e^-$ linear collider with $\sqrt{s} = 300...500$ GeV is ideally suited [4, 5, 6, 7]. Its realization is still uncertain, so it is of major interest whether an intermediate mass Higgs boson can also be found - or excluded - at a future hadron collider like the LHC.

Hadron colliders provide large production rates for the standard model Higgs boson over a wide range of Higgs masses [8, 9, 10, 11] . Because of the large background at hadron colliders much work has already been devoted to the development of special search strategies for different Higgs mass regions [12, 13]. A heavy Higgs boson ( $m_H > 2m_Z$ ) can be found at the LHC quite straightforwardly via the rare decay channel $H \to ZZ \to 4\mu^{\pm}$ up to a Higgs mass of 800 GeV for an integrated luminosity of $10^5$ pb$^{-1}$ [14, 15]. With the help of other processes this range might be extended to 1 TeV [16].

For the mass range $m_Z < m_H < 140$ GeV it is not yet clear whether a standard model Higgs boson can be found at the LHC. The production cross section is large ( $\sigma > 40$ pb ) so that with an integrated luminosity of $10^5$ pb$^{-1}$ per year of running more than $4 \cdot 10^6$ Higgs particles will be produced [9, 17]. In this mass region the Higgs will mainly decay via $H \to b\bar{b}$ and $H \to \tau^+\tau^-$ which are both accompanied by an overwhelming QCD background that makes Higgs searches for these decay channels impossible [18, 19, 20]. The situation is better for the rare decay channel $H \to \gamma\gamma$. It may be successful for the mass range 110 GeV $< m_H <$ 140 GeV [21]. Even for $m_H = 100$ GeV the Higgs boson might be found assuming a dedicated detector. Its precision electromagnetic calorimeter would have to provide a very good energy resolution, a fine granularity and a small time constant [22]. The vertex uncertainty due to the longitudinal extension of the proton bunches would be reduced by using additional information from other parts of the detector besides the calorimeter [23]. However there is a severe jet-jet background where the jets contain single $\pi^0$'s which carry a large fraction of the jet transverse energy. The decay photon pairs of the $\pi^0$'s will give a single electromagnetic energy deposit and will thus fake a photon event [21]. The associated production of H and Z or H and W improves this situation somewhat: e.g. the process $q\bar{q} \to H\ W \to \gamma\ \gamma\ \nu\ \ell$ allows tagging for one isolated lepton and two isolated photons [24, 25, 26, 27]. Another possibility is the associated production of H and $t\bar{t}$ which also leads to the inclusive production of one isolated lepton and two isolated photons  $gg\ (q\bar{q}) \to H\ t\bar{t} \to H\ W + X \to \gamma\ \gamma\ \nu\ \ell + X$ [28, 29]. Finding or excluding a standard model Higgs boson at the LHC with one of these reactions may be achievable for the mass range $m_Z < m_H < 140$ GeV [17, 21, 30, 31] even if the mass resolution requirements demanded for the detector are severe [32].

For the mass range 140 GeV $< m_H < 2m_Z$  a Higgs boson can be found via the process $H \to ZZ \to 4\ell^{\pm}$ with one or both Z bosons being off-shell [12, 19, 33]. The branching ratio for this decay channel is small [9] but it results in an extremely clear signal at a hadron collider. So it is important to decide on how far a Higgs search via this decay mode can be extended to smaller Higgs masses. The problem is that the branching ratio for $H \to ZZ \to 4\ell^{\pm}$ falls steeply for Higgs masses smaller than around 125 GeV because then both Z bosons become off-shell [30]. So one has to know the size of the background. Still missing was so far the irreducible background coming from gluon fusion via a Z boson pair into two charged lepton pairs $gg \to ZZ \to 4\ell^\pm$. We calculate this process and compare it with the irreducible background from quark-antiquark annihilation $q\bar{q} \to ZZ \to 4\ell^\pm$ and with the Higgs signal $gg \to H \to ZZ \to 4\ell^\pm$ in leading order.

The Feynman diagrams for the background from gluon fusion via off-shell Z boson pairs are shown in figure 1. Figure 2 gives the diagrams for the Higgs signal. The gluon fusion background is suppressed in comparison to the quark-antiquark background by a factor $\alpha_s^2$. However this is at least partially compensated by the large gluon luminosity at the low values of Bjorken x involved and by the constructive interference from the different quarks inside the loop. In order to compare, the quark-antiquark background is estimated as in [34]. Not included are contributions containing $\gamma Z$ intermediate states, which were shown to be small [35].
The outline of our paper is as follows. In the next section we investigate the $ggZZ$ polarization tensor. In the third section we give some relations between the corresponding helicity amplitudes. First results from our Monte Carlo phase space calculation are discussed in the fourth section. In section five we give our conclusion.

## 2. The $ggZZ$ polarization tensor

The gluon fusion into two on-shell Z bosons without successive decay into a charged lepton pair has been calculated in [36, 37, 38]. The leptonic decay of the Z bosons was included in [39] using the narrow width approximation where both Z bosons are restricted to the mass shell. We remove that restriction and allow for arbitrary invariant masses. Twelve Feynman diagrams contribute to the background process $gg \to ZZ \to 4\ell^\pm$. The six generic Feynman diagrams for the case that all final state leptons are muons, which are easiest to detect at a hadron collider, are given in figure 1. We keep close to the notation of [38, 39] in order to allow for easy comparison of our results with the on-shell case.
In this section we investigate the gluon fusion via a quark box into two Z bosons with invariant masses $m_3$ and $m_4$. The decay of the two intermediate state Z bosons is included afterwards. We take all four momenta to be outgoing. The two gluons are described by the four momenta $p_1^\mu$, $p_2^\nu$ and the two Z bosons by $p_3^\rho$, $p_4^\sigma$. The gluons are taken on-shell. Therefore

$$p_1^2 = 0, \quad p_2^2 = 0, \quad p_3^2 =: m_3^2, \quad p_4^2 =: m_4^2. \tag{2.1}$$

We use Mandelstam variables

$$s := (p_1 + p_2)^2, \quad u := (p_1 + p_3)^2, \quad t := (p_2 + p_3)^2 \tag{2.2}$$

and

$$\begin{aligned} u_3 &:= u - m_3^2 & u_4 &:= u - m_4^2 \\ t_3 &:= t - m_3^2 & t_4 &:= t - m_4^2 \\ s_3 &:= s - m_3^2 & s_4 &:= s - m_4^2 \end{aligned}$$

$$v := ut - m_3^2 m_4^2 \qquad v_2 := ut - 2m_3^2 m_4^2 \qquad \lambda := (u+t)^2 - 4m_3^2 m_4^2. \tag{2.3}$$

The calculation of the $ggZZ$ polarization tensor is straightforward. We use the Passarino-Veltman scheme [40] for the reduction of the tensor integrals and the 't Hooft-Veltman scheme [41] for the scalar integrals. In intermediate steps we get of the order of $10^5$ terms. So the use of an algebraic manipulation program is mandatory. Throughout our analytical investigations we use FORM by J.A.M. Vermaseren [42]. The reduction of the scalar integrals was checked numerically [43].

As in the on-shell case [38] the $ggZZ$ polarization tensor can be written as

$$P^{\mu\nu\rho\sigma} = -\frac{g_s^2 g_Z^2}{(2\pi)^4} \frac{\delta^{\alpha\beta}}{2} \left\{ v_q^2 P_v^{\mu\nu\rho\sigma} + a_q^2 P_a^{\mu\nu\rho\sigma} \right\} . \qquad (2.4)$$

Here $v_q$ and $a_q$ denote the coefficients of the vector and of the axial part of the Z boson coupling to a quark line, $-ig_Z(v_q + a_q\gamma_5)\gamma_\mu$. $g_s$ and $g_Z$ are the strong and the weak coupling constants. The minus sign in eq. (2.4) comes from the closed fermion loop. $\alpha, \beta$ are the colour indices of the two gluons and $\delta^{\alpha\beta}/2$ comes from the trace over the two colour matrices. Because of charge conjugation invariance no terms proportional to $v_q a_q$ arise.

The tensor $P_v^{\mu\nu\rho\sigma}$ is well known; it is the QED result for photon photon scattering $\gamma\gamma \to \gamma\gamma$ with the two outgoing photons being off-shell. It has been calculated in [44] by using double dispersion relations. We compared the algebraic part of our result for $P_v^{\mu\nu\rho\sigma}$ with [44] and got complete agreement. $P_a^{\mu\nu\rho\sigma}$ is new for the case that the two Z bosons have different invariant masses. It is quite a long expression but as $P_v^{\mu\nu\rho\sigma}$ is well known in the literature, it is sufficient to give the difference $P_v^{\mu\nu\rho\sigma} - P_a^{\mu\nu\rho\sigma}$ which is - as in the on-shell case [38] - much shorter. The difference can be written as

$$P_v^{\mu\nu\rho\sigma} - P_a^{\mu\nu\rho\sigma} = i\pi^2 m^2 \left\{ T^{\mu\nu\rho\sigma} + \mathcal{P}_{12}(T^{\nu\mu\rho\sigma}) + T_1^{\nu\rho\sigma} p_1^\mu + T_2^{\mu\rho\sigma} p_2^\nu \right\} \qquad (2.5)$$

where $m$ is the mass of the quark in the loop. By $\mathcal{P}_{12}$ we denote the permutation that exchanges $p_1$ and $p_2$. Note that this includes the interchange of

$\mathcal{P}_{12}$ :

$$\begin{aligned}
p_1 &\leftrightarrow p_2 & C_0(-p_1, -p_3) &\leftrightarrow C_0(-p_2, -p_3) \\
u &\leftrightarrow t & C_0(-p_2, -p_4) &\leftrightarrow C_0(-p_1, -p_2 - p_3) \\
u_3 &\leftrightarrow t_3 & D_0(-p_1, -p_2, -p_3) &\leftrightarrow D_0(-p_1, -p_2, -p_4) \\
u_4 &\leftrightarrow t_4 & &
\end{aligned}$$

$invariant\ stay$ :

$$s, m_3, m_4, s_3, s_4, C_0(-p_1, -p_2), C_0(-p_3, -p_1 - p_2), D_0(-p_1, -p_3, -p_2) . \qquad (2.6)$$

When calculating the helicity amplitudes in section 3 all terms in eq. (2.5) containing $p_1^\mu$ or $p_2^\nu$ vanish after contraction with the helicity four vectors. So $T_1^{\nu\rho\sigma}$ and $T_2^{\mu\rho\sigma}$ are not needed for the calculation of the helicity amplitudes. It is sufficient to give $T^{\mu\nu\rho\sigma}$

$T^{\mu\nu\rho\sigma} =$

$$16 \left( (\delta^{\mu\sigma} p_1^\rho p_3^\nu - \delta^{\mu\rho} p_1^\sigma p_3^\nu) + \frac{t_3}{s} (\delta^{\mu\rho} p_1^\nu p_1^\sigma - \delta^{\mu\sigma} p_1^\nu p_1^\rho) \right) \left\{ \frac{1}{v} \left( -2s\, C_0(-p_1, -p_2) - 2t_3 C_0(-p_2, -p_3) \right. \right.$$

$$\left. \left. - 2u_4 C_0(-p_2, -p_4) + st\, D_0(-p_1, -p_2, -p_3) + su\, D_0(-p_1, -p_2, -p_4) \right) + D_0(-p_1, -p_3, -p_2) \right\}$$

$$+ 16 \left( \delta^{\mu\sigma} p_1^\nu p_2^\rho + \delta^{\nu\rho} p_1^\sigma p_2^\mu - \delta^{\mu\nu} p_1^\sigma p_2^\rho \right) \left\{ D_0(-p_1, -p_2, -p_3) - D_0(-p_1, -p_2, -p_4) \right.$$

$$\left. - D_0(-p_1, -p_3, -p_2) \right\} - 8 \left( 2 - \frac{\lambda}{v} \right) \left( \delta^{\mu\nu} + \frac{2s}{v} p_3^\mu p_3^\nu \right) \delta^{\rho\sigma} C_0(-p_3, -p_1 - p_2)$$

$$
\begin{aligned}
&+ \frac{16}{v}\left(2 + \frac{u_3 u_4 + t_3 t_4}{v}\right)\left(s p_3^\mu p_3^\nu + m_3^2 p_1^\nu p_2^\mu\right)\delta^{\rho\sigma} C_0(-p_1, -p_2)\\
&+ 8s\frac{u+t}{v}\left(2\frac{u_3 p_2^\mu p_3^\nu + t_3 p_1^\nu p_3^\mu}{v} - \delta^{\mu\nu}\right)\delta^{\rho\sigma} C_0(-p_1, -p_2)\\
&+ 8s\left(\delta^{\mu\rho}\delta^{\nu\sigma} - \delta^{\mu\sigma}\delta^{\nu\rho}\right)D_0(-p_1, -p_2, -p_3) + 4s\left(\delta^{\mu\rho}\delta^{\nu\sigma} + \delta^{\mu\sigma}\delta^{\nu\rho}\right)D_0(-p_1, -p_3, -p_2)\\
&+ 4\delta^{\mu\nu}\delta^{\rho\sigma}\Bigg\{4\frac{u_3 u_4}{sv}\left(u_3 C_0(-p_1, -p_3) + u_4 C_0(-p_2, -p_4)\right)\\
&+ 2\left(2\frac{u^2 s}{v} + 8m^2 - s\right)D_0(-p_1, -p_2, -p_4) + \left(\frac{2v}{s} + 8m^2 - s\right)D_0(-p_1, -p_3, -p_2)\Bigg\}\\
&+ \frac{32}{v}\delta^{\rho\sigma} p_3^\mu p_3^\nu \Bigg\{\left(1 + \frac{u_3 u_4}{v}\right)\left(u_3 C_0(-p_1, -p_3) + u_4 C_0(-p_2, -p_4)\right) + \Big(-v + 2\left(sm^2 - m_3^2 m_4^2\right)\\
&+ \frac{1}{v}\left(u^2\left(t^2 + s^2\right) - m_3^4 m_4^4\right)\Big) D_0(-p_1, -p_2, -p_4) + sm^2 D_0(-p_1, -p_3, -p_2)\Bigg\}\\
&+ 8\delta^{\rho\sigma} p_1^\nu p_2^\mu \Bigg\{4\left(-\frac{m_3^2 u}{v^2} + \frac{1}{s^2}\right)\left(u_3 C_0(-p_1, -p_3) + u_4 C_0(-p_2, -p_4)\right)\\
&+ \frac{2}{v}\left(2m_4^2 - s + \frac{\lambda + 2m_3^2 m_4^2 - m_3^4 - m_4^4}{s} + \frac{m_3^2 \lambda}{v}\right)C_0(-p_3, -p_1 - p_2)\\
&+ 2\left(1 + \frac{2m_3^2 u^2 s}{v^2} + 4m^2\left(\frac{m_3^2}{v} - \frac{1}{s}\right)\right)D_0(-p_1, -p_2, -p_4)\\
&+ \left(1 - \frac{4m^2}{s} - \frac{2v}{s^2}\right)D_0(-p_1, -p_3, -p_2) + \frac{4m^2 m_3^2}{v}D_0(-p_1, -p_3, -p_2)\Bigg\}\\
&+ \frac{16}{v}\delta^{\rho\sigma}\left(u_3 p_2^\mu p_3^\nu + t_3 p_1^\nu p_3^\mu\right)\Bigg\{\frac{2u}{v}\left(u_3 C_0(-p_1, -p_3) + u_4 C_0(-p_2, -p_4)\right)\\
&+ \left(2 - \frac{\lambda}{v}\right)C_0(-p_3, -p_1 - p_2) - 2\left(2m^2 + \frac{su^2}{v}\right)D_0(-p_1, -p_2, -p_4)\\
&- 2m^2 D_0(-p_1, -p_3, -p_2)\Bigg\}
\end{aligned}
\tag{2.7}
$$

where $C_0$, $D_0$ are the scalar three- and four-point functions

$$
C_0(p_1, p_2) :=
$$

$$
\frac{1}{i\pi^2}\int d^4 q \frac{1}{[(q^2 - m^2 + i\epsilon][(q+p_1)^2 - m^2 + i\epsilon][(q+p_1+p_2)^2 - m^2 + i\epsilon]} \tag{2.8}
$$

$$
D_0(p_1, p_2, p_3) := \frac{1}{i\pi^2}\int d^4 q
$$

$$
\frac{1}{[q^2 - m^2 + i\epsilon][(q+p_1)^2 - m^2 + i\epsilon][(q+p_1+p_2)^2 - m^2 + i\epsilon][(q+p_1+p_2+p_3)^2 - m^2 + i\epsilon]}.
\tag{2.9}
$$

This result agrees for $m_3^2 = m_4^2 = m_Z^2$ with the on-shell result in [38]. In addition we checked the gauge invariance of the full polarization tensor $P^{\mu\nu\rho\sigma}$. Gauge invariance leads to

$$
p_{1\mu} P^{\mu\nu\rho\sigma} = p_{2\nu} P^{\mu\nu\rho\sigma} = 0
$$

$$p_{3\rho} P_v^{\mu\nu\rho\sigma} = p_{4\sigma} P_v^{\mu\nu\rho\sigma} = 0$$
$$for \quad m = 0: \qquad p_{3\rho} P_a^{\mu\nu\rho\sigma} = p_{4\sigma} P_a^{\mu\nu\rho\sigma} = 0 \qquad (2.10)$$

where m is again the mass of the quark in the loop.

## 3. The $ggZZ$ helicity amplitudes

In order to get the helicity amplitudes from the full ggZZ polarization tensor in eq. (2.4) we specify explicitly a set of four momenta $p_1^\mu, p_2^\nu, p_3^\rho, p_4^\sigma$ for the two gluons and the two Z bosons as well as a corresponding set of helicity four vectors. We use in the parton center of momentum frame

$$\begin{aligned}
p_1^\mu &:= (-p, 0, 0, -p) \\
p_2^\nu &:= (-p, 0, 0, +p) \\
p_3^\rho &:= (P_3, q\sin\theta, 0, q\cos\theta) \\
p_4^\sigma &:= (P_4, -q\sin\theta, 0, -q\cos\theta)
\end{aligned} \qquad (3.1)$$

with

$$\begin{aligned}
P_3 &= \frac{1}{2\sqrt{s}}(s + m_3^2 - m_4^2) \\
P_4 &= \frac{1}{2\sqrt{s}}(s + m_4^2 - m_3^2) \\
\sin\theta &= 2\frac{\sqrt{ut - m_3^2 m_4^2}}{\sqrt{\lambda}} \\
p &= \frac{\sqrt{s}}{2} \quad, \quad q = \frac{\sqrt{\lambda}}{2\sqrt{s}} \, .
\end{aligned} \qquad (3.2)$$

For the helicity four vectors we choose

$$\begin{aligned}
e_1^{+\mu} &= -e_2^{-\mu} = \frac{1}{\sqrt{2}}(0, -i, 1, 0) \\
e_1^{-\mu} &= -e_2^{+\mu} = \frac{1}{\sqrt{2}}(0, +i, 1, 0) \\
e_3^{+*\mu} &= -e_4^{-*\mu} = \frac{1}{\sqrt{2}}(0, +i\cos\theta, 1, -i\sin\theta) \\
e_3^{-*\mu} &= -e_4^{+*\mu} = \frac{1}{\sqrt{2}}(0, -i\cos\theta, 1, +i\sin\theta) \\
e_3^{0*\mu} &= \frac{1}{m_3}(q, +P_3\sin\theta, 0, +P_3\cos\theta) \\
e_4^{0*\mu} &= \frac{1}{m_4}(q, -P_4\sin\theta, 0, -P_4\cos\theta)
\end{aligned} \qquad (3.3)$$

where $+, -, 0$ refer to positive, negative and longitudinal polarization respectively. We get $2^2 * 3^2 = 36$ helicity amplitudes. They are related by parity

$$\mathcal{M}_{\lambda_1 \lambda_2 \lambda_3 \lambda_4} = (-1)^{\lambda_3 + \lambda_4} \mathcal{M}_{-\lambda_1 -\lambda_2 -\lambda_3 -\lambda_4} \, . \qquad (3.4)$$

Moreover we find the relations

$$\begin{aligned}
\mathcal{M}_{++--}(\sqrt{\lambda}) &= \mathcal{M}_{++++}(-\sqrt{\lambda}) \\
\mathcal{M}_{+++-}(\sqrt{\lambda}) &= \mathcal{M}_{++-+}(-\sqrt{\lambda}) \\
\mathcal{M}_{+---}(\sqrt{\lambda}) &= \mathcal{M}_{+-++}(-\sqrt{\lambda}) \\
\mathcal{M}_{+--+}(\sqrt{\lambda}) &= \mathcal{M}_{+-+-}(-\sqrt{\lambda}) \\
\mathcal{M}_{++0+}(\sqrt{\lambda}) &= -\mathcal{M}_{++0-}(-\sqrt{\lambda}) \\
\mathcal{M}_{+++0}(\sqrt{\lambda}) &= -\mathcal{M}_{++-0}(-\sqrt{\lambda}) \\
\mathcal{M}_{+-+0}(\sqrt{\lambda}) &= -\mathcal{M}_{+--0}(-\sqrt{\lambda}) \\
\mathcal{M}_{+-0-}(\sqrt{\lambda}) &= -\mathcal{M}_{+-0+}(-\sqrt{\lambda}) \ .
\end{aligned} \qquad (3.5)$$

So only ten helicity amplitudes are independent. We choose the following set of independent amplitudes

$$\begin{array}{ccccc}
\mathcal{M}_{++++} & \mathcal{M}_{+++0} & \mathcal{M}_{+++-} & \mathcal{M}_{++0+} & \mathcal{M}_{++00} \\
\mathcal{M}_{+-++} & \mathcal{M}_{+-+0} & \mathcal{M}_{+-+-} & \mathcal{M}_{+-0+} & \mathcal{M}_{+-00}
\end{array} \qquad (3.6)$$

We checked that all our helicity amplitudes are ultraviolet finite. Moreover they agree for $m_3^2 = m_4^2 = m_Z^2$ with the results from [38]. We were able to bring our analytical result for eq. (3.6) into a relatively compact form so that all ten helicity amplitudes together contain only some 4700 terms. This eases the numerical evaluation considerably.

## 4. Numerical results

In order to numerically evaluate our analytical result for eq. (3.6) we let each intermediate state Z boson decay into a charged muon pair and integrate the partonic cross section $\hat{\sigma}$ over the gluon structure functions of the two incoming protons

$$\sigma(pp \to gg \to ZZ \to 4\mu^\pm) = \int dx_1 dx_2 f_g(x_1, Q^2) f_g(x_2, Q^2) \hat{\sigma}(\hat{s}) \ . \qquad (4.1)$$

Here $\hat{s} = x_1 x_2 s$ with $s$ being the C.M. energy squared of the hadrons. $x_1, x_2$ are the Bjorken x variables of the two gluons, $f_g$ is the gluon structure function of the proton and $Q^2$ determines the scale to which the structure functions are evolved. We use the updated set D_ by Martin, Roberts and Stirling which includes the NMC deep inelastic scattering data [45, 46, 47]. For the QCD scale we take $\Lambda_{4\text{flavours}} = 215$ MeV and for $Q^2$ we choose $Q^2 = \hat{s}/4$. The energy range under consideration lies between the bottom quark and the top quark threshold. So we use for the strong coupling constant $\alpha_s = 12\pi/23 \log(Q^2/\Lambda^2)$ with $\Lambda = \Lambda_{5\text{flavours}} = 143$ MeV [48]. We get the partonic cross section $\hat{\sigma}$ by integrating over the four particle final state. All final state muons are taken to be massless. In accordance with eq. (2.1) and with figure 1 we denote the three momenta by

$$\begin{array}{ll}
\vec{p}_a, \vec{p}_c & \text{for the two final state } \mu^- \\
\vec{p}_3, \vec{p}_4 & \text{for the two intermediate state Z bosons} \ .
\end{array} \qquad (4.2)$$

Then we get for the partonic cross section [49]

$$\hat{\sigma}(\hat{s}) = \frac{1}{2\hat{s} \cdot 8^3 \hat{s} \cdot (2\pi)^8} \int dm_3^2 dm_4^2 d\Omega_3 d\Omega_a d\Omega_c \sqrt{\lambda(\hat{s}, m_3^2, m_4^2)} \mid \mathcal{M} \mid^2 \qquad (4.3)$$

with $\lambda(\hat{s}, m_3^2, m_4^2) = \lambda$ from eq. (2.3). $\int d\Omega$ stands for the integration over the solid angle of the corresponding three vector. $\mathcal{M}$ is the sum over all interfering invariant subamplitudes that contribute to the given overall process. So it contains our analytical result eq. (3.6) for the ggZZ part of a Feynman diagram, both Z propagators and the lepton currents that describe the decay of the two Z bosons. From eqs. (4.1) and (4.3) it is clear that we end up with a ten dimensional integral where only the integration over the azimuthal angle around the beam axis is trivial. The remaining nine integrations are done numerically with a selfadaptive Monte Carlo integration routine [50].

We include three processes into our numerical investigation

$$\text{the signal} \qquad pp \to gg \to \quad H \to ZZ \to 4\mu^{\pm} \qquad (4.4)$$
$$\text{the two irreducible} \qquad pp \to gg \to \quad \text{quarkbox} \to ZZ \to 4\mu^{\pm} \qquad (4.5)$$
$$\text{backgrounds} \qquad pp \to q\bar{q} \to \quad ZZ \to 4\mu^{\pm} \ . \qquad (4.6)$$

For all these processes we restrict ourselves to the leading order contributions. In our analysis we use the LHC design values for the energy in the proton rest frame ($\sqrt{s} = 14$ TeV) and for the integrated luminosity per year of running ($\mathcal{L} = 10^2$ fb$^{-1}$). We assume $m_{\text{top}} = 140$ GeV. For the other quarks in the loop we take the massless limit. In order to avoid numerical problems we make a cut on the transverse momentum of each Z boson $\mid \vec{p}_{T\,Z} \mid > 2$ GeV. The contribution from the hereby neglected part of the phase space is small. For a crude simulation of detector properties we use cuts on the transverse momentum and on the rapidity along the beam axis for each final state lepton :  $\mid \vec{p}_{T\,\ell\text{epton}} \mid > 10$ GeV and  $\mid y_{z\,\ell\text{epton}} \mid < 2.5$ [51, 52]. The total width of the Higgs boson we take from [53] using a running b quark mass $m_b(4.89 \text{ GeV}) = 4.89$ GeV   [54].

We checked our numerical program thoroughly. When restricting the intermediate state Z bosons to the mass shell we got perfect agreement with the numerical results from [39] for all three reactions (4.4), (4.5) and (4.6). Comparison with [38] also gave good agreement when using two on-shell Z bosons as the final state. The background (4.6) from quark-antiquark annihilation with two on-shell Z bosons as a final state is also in complete numerical agreement with the leading order results from [55] . As an off-shell test we checked that the cross section for the signal (4.4) gives in the limit of a very heavy top quark the correct value.

In figure 3 we show the invariant mass ($m_{ZZ}$) distribution of the Z Z intermediate state for the signal (4.4) and for the two irreducible background processes (4.5) and (4.6). $m_{ZZ}$ is identical to the invariant mass of the four lepton final state. In figure 3 all final state leptons are muons. The differential cross sections for the two background reactions fall rapidly for invariant masses below the Z Z threshold. The contribution from gluon fusion is always smaller than the contribution from quark-antiquark annihilation. The relative size of these two differential cross sections is given in figure 4. Here each value is averaged over a 10 GeV invariant mass bin width in order to account for the finite energy resolution of the detector. The gluon fusion contribution is around 20 % of the contribution from quark-antiquark annihilation. The bars in figure 3 denote the cross section per GeV for the signal (4.4) averaged over a 10 GeV invariant mass bin width. The Higgs mass lies at the center of each bin. The numbers on top of each bar give the number of signal events per 10 GeV bin width and per year of running. The signal falls rapidly for small invariant masses due to the vanishing branching ratio for H → Z Z with one and finally both Z bosons being off-shell. The signal shows around 170 GeV the well known characteristic dip. It results from the opening of the decay channel H → W$^+$ W$^-$ with both W's being on-shell. Figure 3 shows clearly that in the lower half of the intermediate mass region ($m_Z < m_H < 2\, m_Z$) the irreducible background from gluon fusion and quark-antiquark annihilation via off-shell Z boson pairs is negligible.

Instead of four muons in the final state one can also search for a muon pair accompanied by an electron pair. For this process only half of the Feynman diagrams in figures 1 and 2 contribute. Moreover the statistical factor $1/2! \cdot 2!$ accounting for two pairs of identical particles does not appear. So one would naively expect the cross section for the $2e^\pm\, 2\mu^\pm$ final state to be bigger by roughly a factor 2 than the cross section for the $4\mu^\pm$ final state if the interference between the permuted and the non permuted $4\mu^\pm$ diagrams is small. This is clearly fullfilled when both Z bosons are on-shell because then the permutation of two identical muons in the final state would lead in most cases to an event with one or even both intermediate state Z bosons being off-shell. So the permuted event would then be suppressed by the Z propagator. Figure 5 confirms this expectation. It shows the ratio of the cross sections for the signal for the $4\mu^\pm$ and for the $2e^\pm\, 2\mu^\pm$ final state. Each plotted value is averaged over a 10 GeV invariant mass bin width. Above the Z Z threshold the ratio is within the statistical Monte Carlo errors indeed equal to two. At smaller values of the invariant mass the $4\mu^\pm$ final state shows a clear constructive interference. This enhances the signal (4.4) for Higgs masses below 120 GeV increasingly. However the event rates for the signal are already too small to be seen due to the vanishing branching ratio for $H \to ZZ$. Looking at the number of signal events given in figure 3 one expects that the intermediate mass Higgs search via off-shell Z boson pairs at the LHC will be successful down to Higgs masses of about 130 GeV.

## 5. Conclusion

We investigated the size of the irreducible background to the intermediate mass Higgs search via off-shell Z boson pairs at hadron colliders. We restricted to those contributions that contain a $Z\,Z$ intermediate state. We found that in the lower half of the intermediate mass region the background from gluon fusion and the background from quark-antiquark annihilation is not severe. The Higgs signal in the region below 120 GeV Higgs mass is increasingly enhanced for the $4\mu^\pm$ final state due to constructive interference between diagrams that differ by the permutation of identical muons in the final state. For the Higgs search this has no relevance because signal rates are already too small in that mass region. We conclude that the search for an intermediate mass standard model Higgs boson via the decay channel $H \to ZZ \to 4\ell^\pm$ at the LHC will be successful down to Higgs masses of about 130 GeV. More problematic seems to be the reducible background [33]. Its main sources are the non-resonant background from $t\bar{t}$ production and the semi-resonant background from $Z b\bar{b}$ production. Here further work is needed for definitive conclusions.

**Acknowledgements** We thank G.Kramer for carefully reading the manuscript.
C.Z. would like to thank A.Ali, G.Jikia, B.A.Kniehl, T.Ohl, H.Simma, M.Spira, J.A.M.Vermaseren and P.M.Zerwas for useful discussions. The use of the DESY HP workstation cluster is gratefully acknowledged.

# Figure Captions

**Fig.1** Feynman diagrams for the background process $gg \to ZZ \to 4\mu^{\pm}$. Both intermediate state Z bosons are allowed to be off-shell.

**Fig.2** Feynman diagrams for the signal process $gg \to H \to ZZ \to 4\mu^{\pm}$. Again both intermediate state Z bosons are allowed to be off-shell.

**Fig.3** The invariant mass ($m_{ZZ}$) distributions for the two irreducible background processes $pp \to q\bar{q} \to ZZ \to 4\mu^{\pm}$ and $pp \to gg \to ZZ \to 4\mu^{\pm}$ (continous $q\bar{q}$ and $gg$ lines). The bars denote the cross section per GeV for the signal $pp \to gg \to H \to ZZ \to 4\mu^{\pm}$ averaged over a 10 GeV invariant mass bin width. The Higgs mass lies at the center of each bin. The numbers on top of each bar give the number of events per 10 GeV bin width and per year of running. ( Integrated luminosity : $10^2 fb^{-1}$, $m_{top} = 140$ GeV, MRS $D_-$ structure functions, cuts as given in the text.)

**Fig.4** Ratio of the cross sections for the two irreducible background contributions $pp \to gg \to ZZ \to 4\mu^{\pm}$ and $pp \to q\bar{q} \to ZZ \to 4\mu^{\pm}$. Each plotted value is the average over a 10 GeV bin width of the invariant mass $m_{ZZ}$. ( $m_{top} = 140$ GeV, MRS $D_-$ structure functions, cuts as given in the text.)

**Fig.5** Ratio of the cross sections for the signal $pp \to gg \to H \to ZZ \to 2\ell^{\pm} 2\ell'^{\pm}$ for the $4\mu^{\pm}$ and for the $2e^{\pm} 2\mu^{\pm}$ final state. Each plotted value is the average over a 10 GeV bin width of the invariant mass $m_{ZZ}$. ( $m_{top} = 140$ GeV, MRS $D_-$ structure functions, cuts as given in the text.)

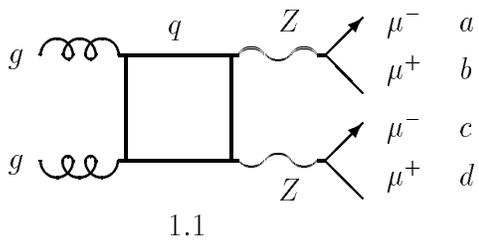

1.1

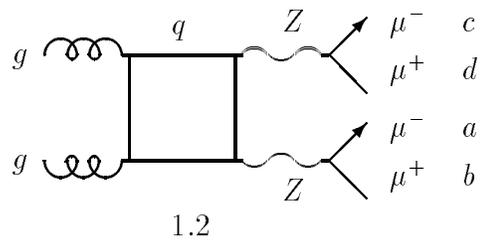

1.2

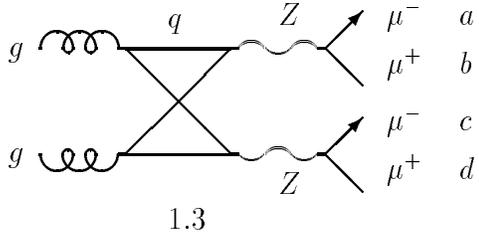

1.3

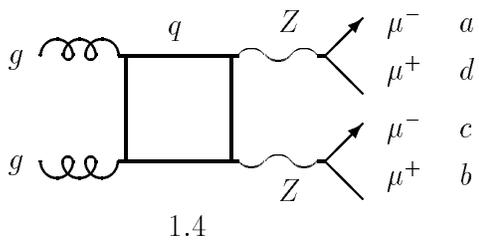

1.4

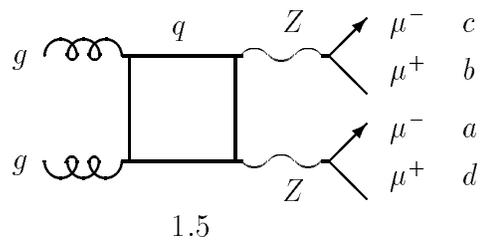

1.5

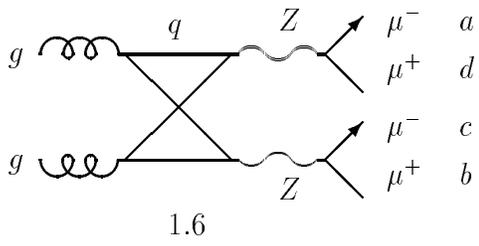

1.6

**Fig.1**

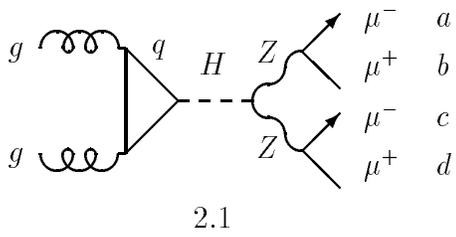

2.1

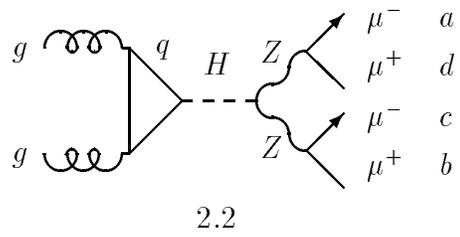

2.2

**Fig.2**

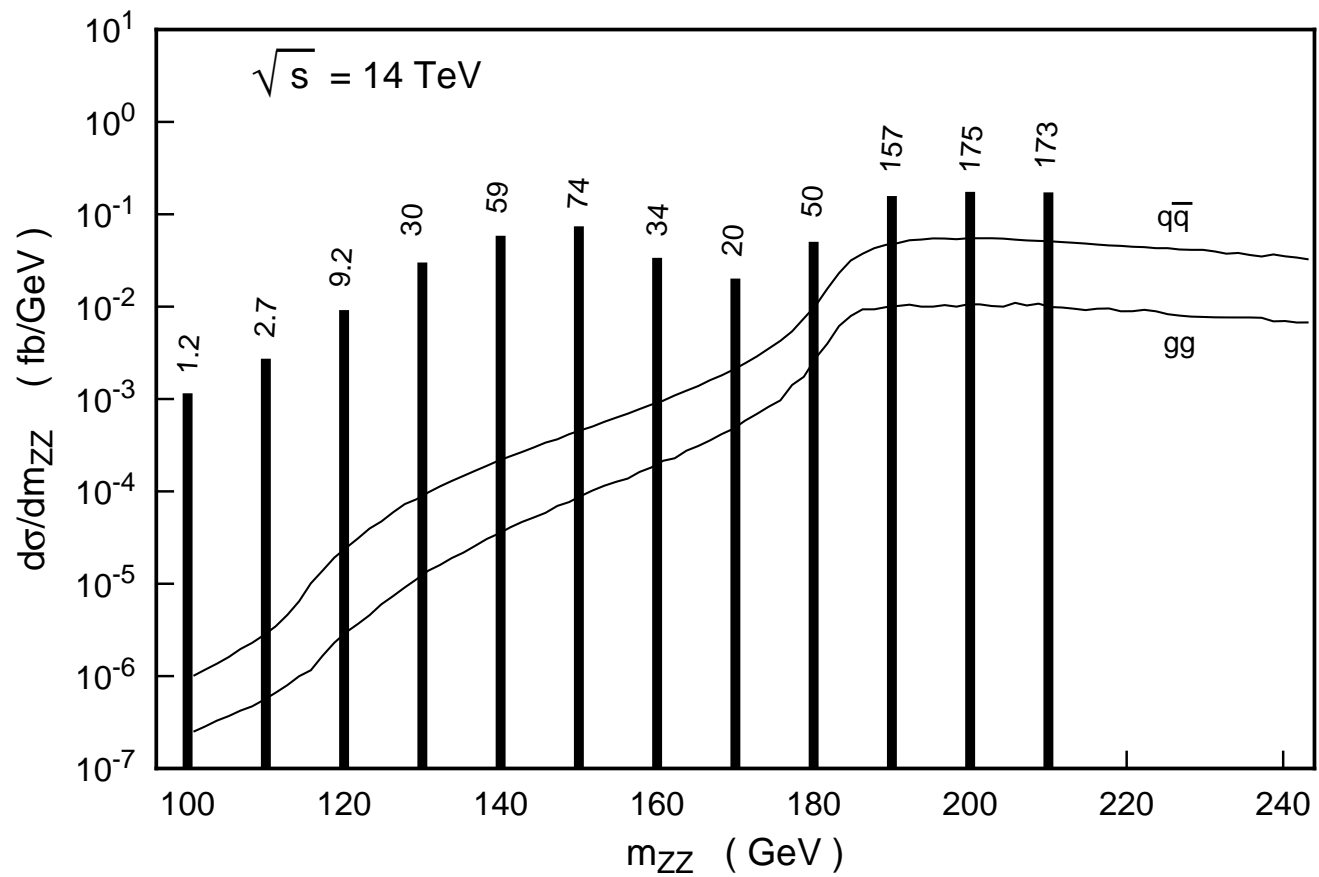

Fig.3

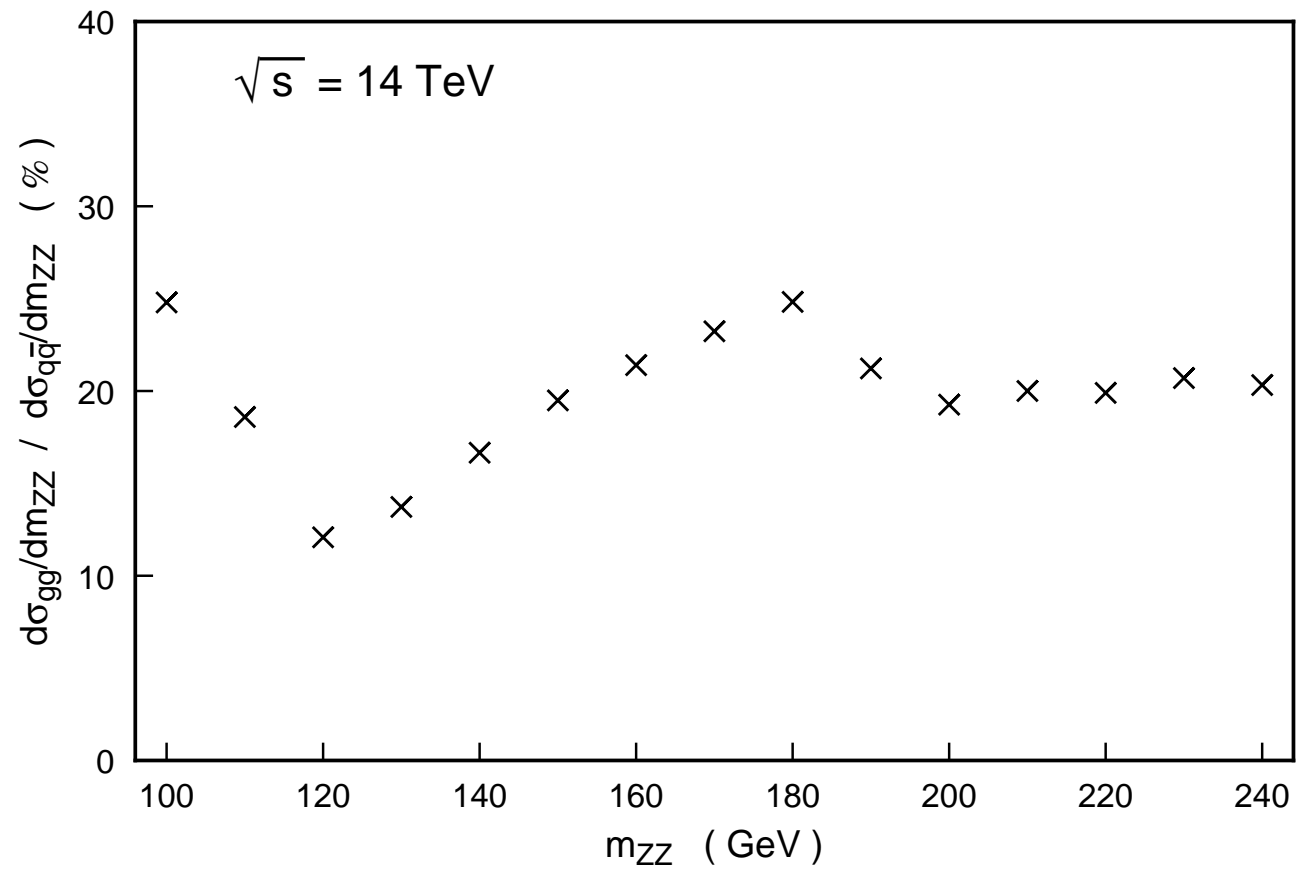

Fig.4

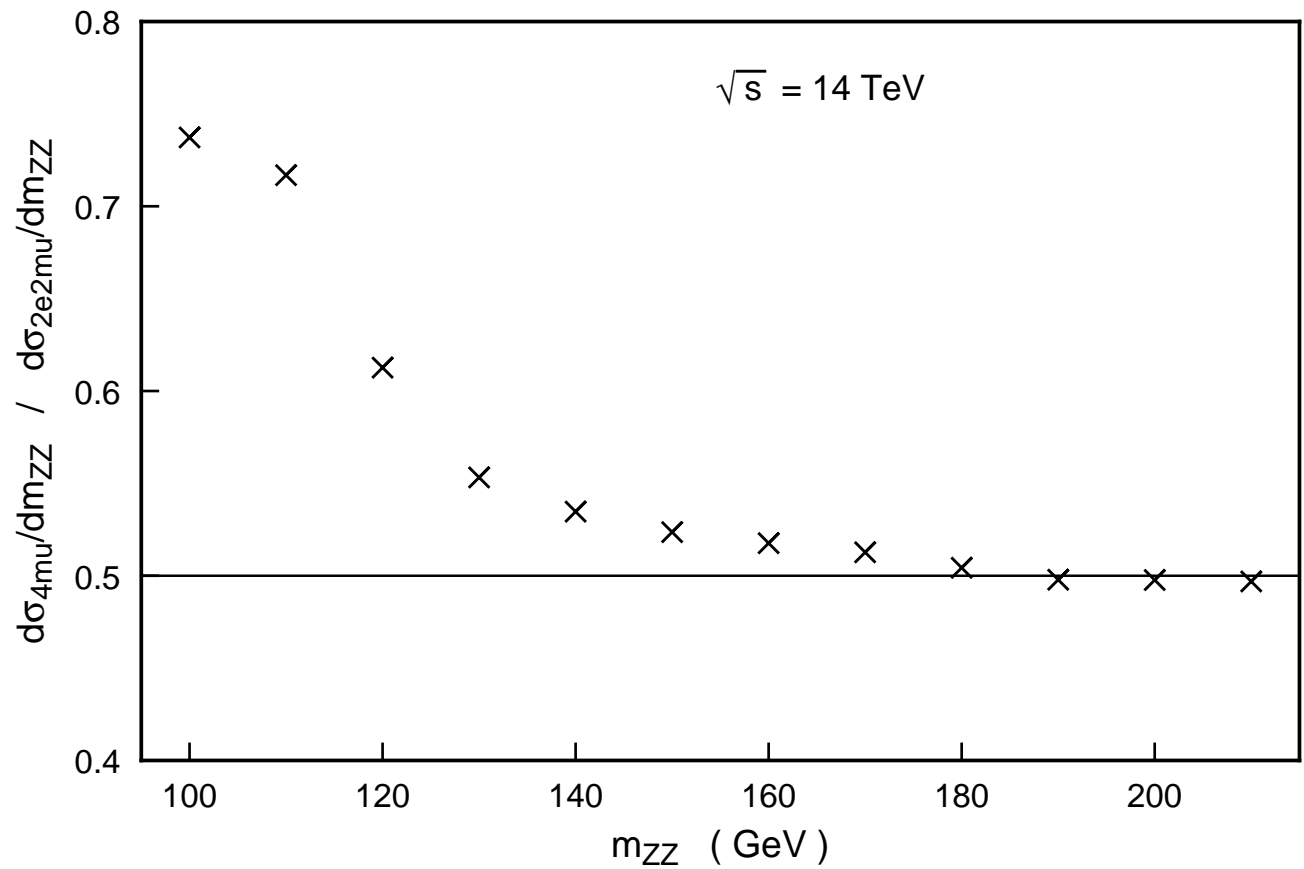

Fig.5